\documentstyle[prl,epsfig,aps,multicol]{revtex}
\begin{document}
\tightenlines

\title {Coexistence of spanning clusters in
directed percolation  }

\author  {Parongama Sen\cite{eml1}}

\address {Department of Physics,
Surendranath College,
24/2 M. G. Road,
Calcutta 700009, India}

\author{Somendra M. Bhattacharjee\cite{eml2}}
\address{Institute of Physics, Bhubaneswar 751 005, India}

\date{\today}
\maketitle

\begin{abstract}
The probability distribution for the number of top to bottom spanning
clusters in Directed percolation in two and three dimensions appears
to be universal and is of the  
form $P(n) \sim \exp(-\alpha n^2 )$.  We argue that $\alpha$ is a new
critical quantity vanishing at the upper critical dimension.  The
probability distribution of the individual masses of the spanning
clusters is found to have a Pearson distribution with a lower
cutoff.  Various properties of the clusters are reported.
\end{abstract}

\begin{multicols}{2}
In a large variety of equilibrium or nonequilibrium critical systems
spatial or space-time anisotropy plays a crucial role. Many such
athermal cases exhibit a continuous transition in the geometric
properties like the sizes or shapes of clusters, and the critical
behavior is best described as belonging to the universality class of
{\it directed percolation} (DP)\cite{dprev}.  Examples abound \cite{regge} as fluid
flow through porous medium in an external field, threshold
dynamics in random systems, forest-fire or epidemic
models, reaction diffusion systems, self
organized criticality, resistance(or insulator)-diode
network, damage spreading, Reggeon field
theory etc.  Naturally, the detailed properties right at
the percolation threshold $p_c$ are influenced by the nature of the
spanning clusters. But, then, how many spanning clusters does one
expect at the threshold?

Let us consider a DP problem on a square or a cubic lattice where each
lattice site is occupied independently with a probability $p$.  
We want percolation that involves a
top to bottom spanning (TB-spanning) cluster. Above the percolation
threshold, there is only one infinite cluster occupying a finite
density of sites of the lattice.  However, at the threshold, a
spanning cluster is a zero-density self-affine fractal (requiring
direction dependent length rescaling).  The lattice can then in
principle support many such TB-spanning clusters.

Although  renormalization group arguments have led to 
 a coherent understanding of long distance (and time)
properties of DP, especially the universal exponents, 
detailed questions like the number of spanning clusters at the
critical point still remains unanswered.  
The known exponents do tell us
that there is a problem in high enough dimensions due to violations of
the hyperscaling relation and this can be
resolved\cite{conglio,dearcangelis} if there is an {\it infinite}
number of spanning clusters at $p=p_c$. This argument is not
applicable for dimensions below the upper critical dimension ($d_c=5$)
and therefore the question stays whether there will be a {\it finite
number} of spanning clusters or a {\it finite density} or {\it just
one}.

In a number of recent
studies\cite{dearcangelis,aizenman,hulin,sen1,schur,sen2,sen3,cardy}, it has
been found that there exists more than one TB-spanning cluster at the
percolation threshold for $d< 6$ for ordinary (i.e., undirected or
isotropic) percolation (OP)\cite{stauffera}.  For two dimensions,
it has  been shown
analytically that there exists more than one TB-spanning cluster at
the percolation threshold and   the
probability $P(n)$ of the existence of $n$ such clusters 
follows  a behavior like $\exp (-\alpha n^\gamma )$ with
$\gamma = 2$ for large $n$ \cite{aizenman}.  
There is some controversy about the
behavior of $P(n)$ in higher dimensions, though recent results
indicate that $\gamma $ is independent of dimension and remains equal
to 2 while $\alpha $ decreases with dimension\cite{sen2}. 
The
individual masses of the spanning clusters have  been shown to
follow the scaling behavior\cite{exponent} of the unique spanning
cluster in all dimensions 2 to 5. It has been conjectured that in two
dimensions, the ratio of the masses of the largest and the second
largest spanning cluster is in all probability universal\cite{sen3}.

In sharp contrast to ordinary percolation\cite{staufferb}, 
very few exact
or rigorous results are available (to the best of our
knowledge) for DP (except mean field results). In order to develop an
analytical understanding, attempts have recently been made to bridge
the gap by studying crossover from OP to DP\cite{opdp}.  Hence the
necessity to know better the difference or similarity between OP and
DP in features that are universal but yet to be resolved in the RG
framework.

Here we have studied coexisting TB-spanning clusters by simulating DP
on square and cubic lattices with helical  boundary conditions in the
transverse directions.  The clusters are identified by using the
Hoshen Kopelman algorithm, respecting the directedness of the
paths.  For any finite lattice, we have first 
estimated the percolation threshold using the method described
in\cite{sen1}. Estimates of $p_c$ in the thermodynamic limit are quite
well-established for DP; however, we have worked with our estimated
$p_c$ whenever finite size effects are strong, especially in three
dimensions and rectangular geometries in two dimensions.  The shape of
the lattice (i.e, when the size of the lattice is not equal in
different dimensions) itself may also play an important role in
determining, e.g., the spanning probabilities etc. Even in OP, it was
shown\cite{hulin} that the TB-spanning probabilities show scaling
behavior with respect to the aspect ratio.  However, the percolation
threshold is not affected except showing stronger finite size effects.
With this in mind, and angular dependence being special to DP\cite{domany},
several variations have been introduced in two dimensions, e.g, two
different directions of propagation have been considered and also the
aspect ratio has been varied.  In three dimensions, we only consider
the DP to be propagating along the diagonal.

Our main results can be summarized as follows:\\ 
(1) We  measure
$P(n)$ the probability for $n$ TB-spanning clusters. From the absence
of any significant finite size effect, we conclude that there is a
nonzero probability of a {\it finite} number of such clusters at
$p=p_c$ both in two and three dimensions.\\ 
(2) The probability
distribution $P(n) \sim \exp(-\alpha n^2)$ and the functional form is
perhaps universal.\\ 
(3) We conjecture that $\alpha$ is a weakly
universal quantity that vanishes above the upper critical dimension
and $\alpha=O(\epsilon)$ for $\epsilon=d_c-d \rightarrow 0$.
This $\alpha$ is a new characteristic for percolation.\\ 
(4) The fractal dimension of the spanning clusters is the same as
expected for 
the unique  cluster at $p_c$.\\
(5) The probability distribution of the masses (from all
configurations) has a universal asymmetric
Pearson distribution (Type III)  in terms of
a scaled mass variable. See Eq. \ref{eq:qx}.\\
(6) The mass ratio of the two largest clusters 
also has a universal probability distribution.\\ 
(7) The ratio of the average masses, 
universal for OP, may have an angular
dependence.

\begin{figure}
{}\hskip -.5in\psfig{file=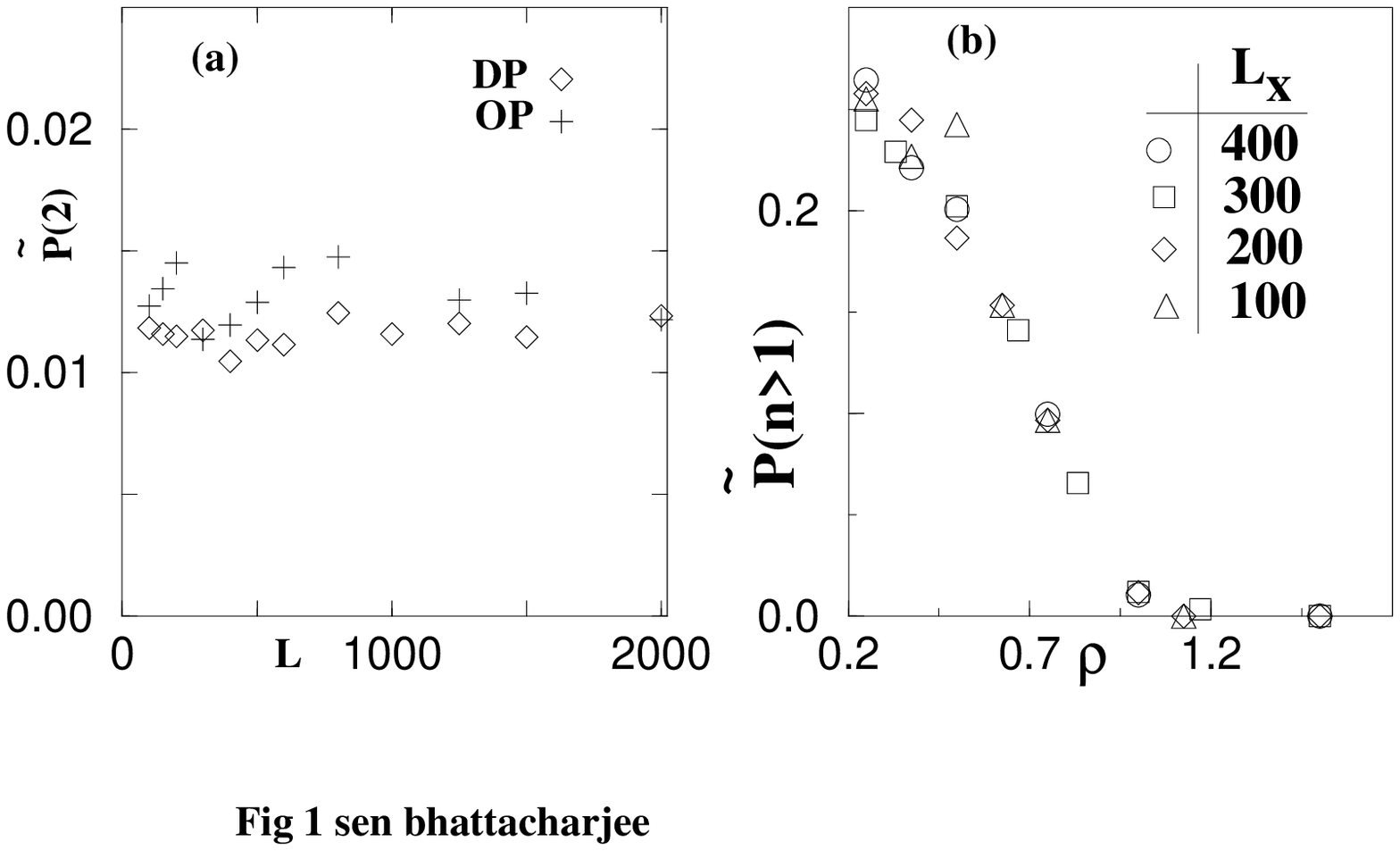,width=4in}
\narrowtext
\caption{ (a) The normalized probability of the existence of two 
spanning clusters for ordinary and directed percolation are
shown for square lattices. The direction of the DP is along the
diagonal and $\tilde P(n > 2) = 0$  for both cases.
(b)The normalized spanning probabilities for $n>2$ for the 
the diagonal DP in two dimensions shown against the aspect ratio.
\label{fig:pvsl}}
\psfig{file=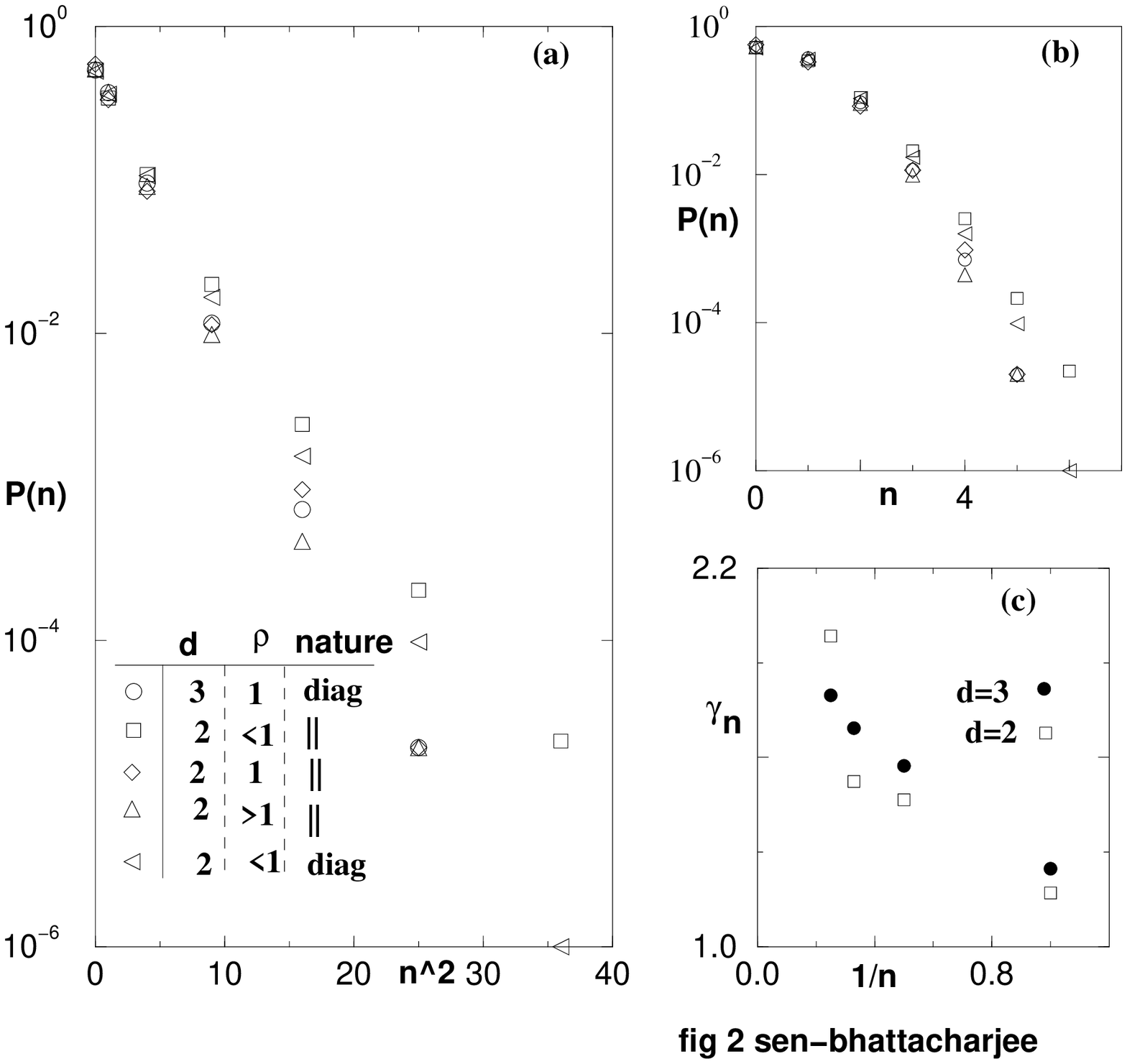,width=3in}
\caption{(a) The probability distribution of spanning clusters plotted
against $n^2$ for different DP's.
(b) Same as (a) but now plotted against $n$.
(c) The plot of $\gamma_n$ vs $1/n$.\label{fig:pvsn2} }
\psfig{file=sb_fig3.eps,width=3in}
\caption{ The masses of different spanning clusters for
DP are shown along with specific power law variations for
comparison. For two dimensions, $L = L_x$ and system sizes
are enlarged by 10 for three dimensions.\label{fig:mvsl} } 
\psfig{file=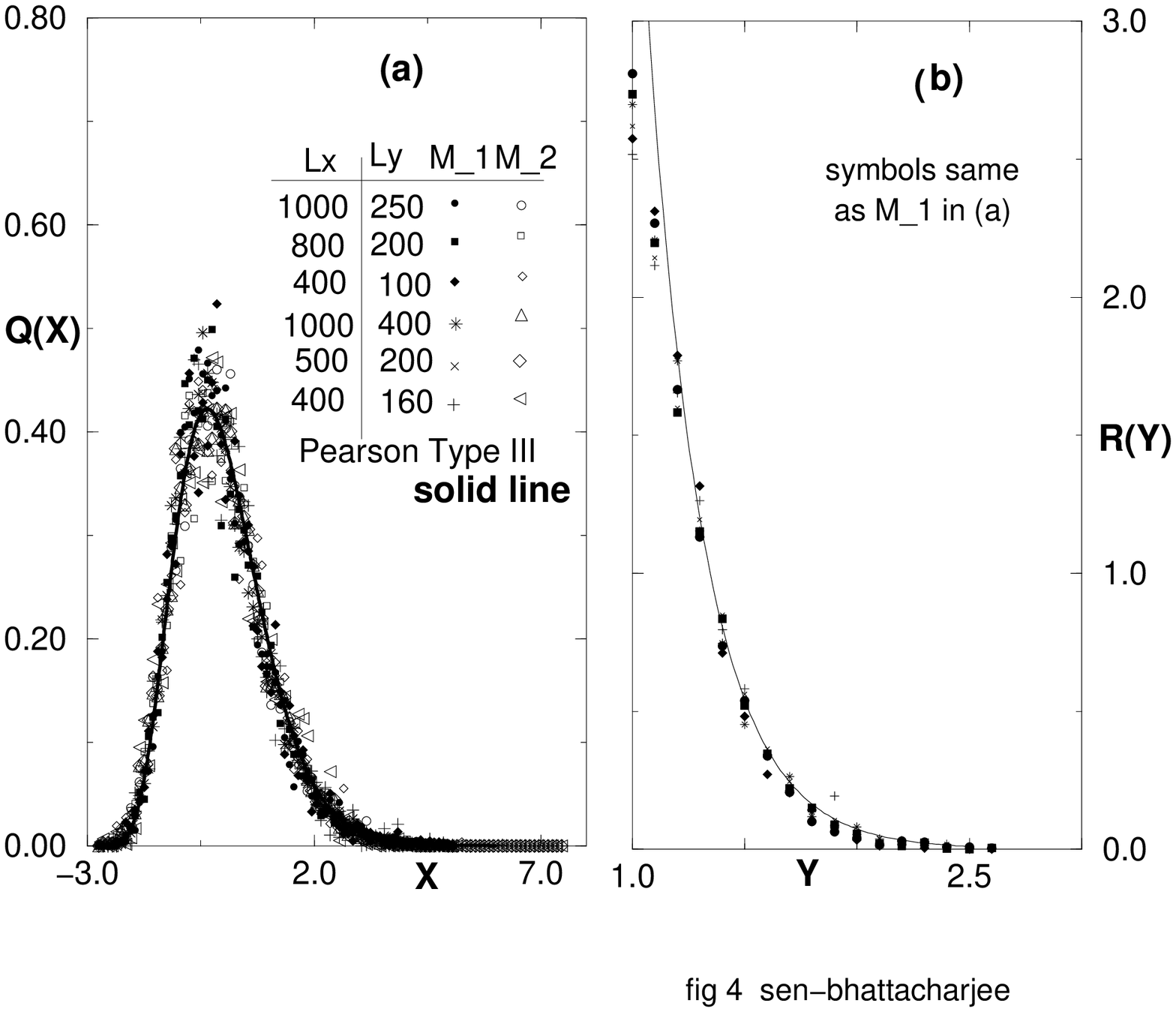,width=3in}
\caption{(a) The probability distribution of masses of individual clusters
irrespective of the total number.  The solid line is the Pearson
distribution type III with zero mean and unit standard deviation with
a lower cutoff $-2.8$. 
(b)The probability distribution for the mass ratio of the two largest
clusters for $n \geq 2$.  The solid line is an exponential distribution
with $A=4.$ \label{fig:pmvsx}}
\end{figure}

In two dimensions, we consider lattices of dimensions $L_x \times
L_y$. With $L_x = L_y$ (square lattices), we check that for both
directions, there is a non-vanishing probability for the existence of
more than one spanning cluster. In the case of the diagonal DP, the
number of spanning clusters $n \leq 2 $ always, and the normalized
probability $\tilde P(2) = {P(2)\over {1-P(0)}}$ here is found to be
not significantly different from that of OP, where the occurrence of
coexisting spanning clusters is quite well established.  This is shown
in Fig \ref{fig:pvsl}a and to be noted is the absence of any
significant finite size effect. 

For the DP parallel to the $y$ axis, a larger number of spanning
clusters is obtained, giving us an opportunity to study the behavior
of $P(n)$.  Since in OP, $P(n) \sim \exp(-\alpha n^\gamma )$ we check
whether this behavior is still true for DP. 
We plot, in Fig \ref{fig:pvsn2},
$P(n)$ against $n$ and $n^2$ and the latter gives a better straight
line, at least for the range of $n$ in the present study.
Fig \ref{fig:pvsn2}c shows a plot of the
finite log-derivative defined as $\gamma_n = \ln(z)/\ln(n/n+1) $ where
$z = \ln(P(n+1))/\ln(P(n))$ for two cases with 
$n \geq 5$ for both $d=2$ and $3$.
This plot of $\gamma_n$ shown against $1/n$, though not very
systematic, is indicative of an extrapolated value of 2 for $n
\rightarrow \infty$ and definitely rules out a small value like
$\gamma=1$.
That we obtain a larger number of clusters, compared to OP,
may be due to a lesser value of $\alpha $ in DP.
Its significance is discussed later. 

For rectangular geometries, i.e., when $L_x \neq L_y$, we
find that for the diagonal DP, $P(n > 1)$ vanishes for 
$\rho = L_y/L_x > 1$. The plot $P(n>1)$ against $\rho$ in Fig
\ref{fig:pvsl}
for different
sets of $L_x, L_y$ shows that it is only dependent on the aspect 
ratio $\rho$, a result which was also obtained in OP\cite{hulin}. A
larger number of  spanning clusters is obtained even for the diagonal
DP when $\rho < 1$. 
No such crossover is observed for the DP
parallel to the spanning direction in the sense $P(n> 1)$ is always
non-zero even for $\rho$ as high as 8.  
These results tell us  that such a crossover behavior 
is dependent on the direction of the DP, apparently vanishing when it is
parallel to the spanning  direction. 
In fact, the crossover in the diagonal (and
not in the parallel) DP is possibly because of the relatively 
larger spread of the clusters in the $x$ direction. 
The study of the probability 
distribution $P(n)$ is  possible for the diagonal DP with $\rho < 1$ as
we get around 5-6 spanning clusters, as well as for the parallel DP 
for any aspect ratio.  We again get results compatible to the 
value  $\gamma = 2$ for each individual case. 

Another interesting behavior is observed for the DP propagating along
the diagonal direction when we consider $\rho < 1$. 
Here $p_c$ decreases as $L_x$ is made larger, 
keeping $L_y$ constant.  
With $L_x/L_y=m$ an integer, one may visualize the
lattice as comprising of $m$ square cells placed adjacent to each
other. We see that though the value of
$p_c$ is higher in each individual cell, it is {\it lesser} for the
entire system. This could happen if the clusters in the individual
cells do {\it not} TB-span but join with those in the neighboring
cells to span the larger lattice.  If this picture 
is correct then the width of the
spanning cluster 
along the $x$ direction should increase with   $L_x$.  We have
verified that this is true.

The mass of the spanning cluster in a DP is expected to behave as
$\langle M_{i,n} \rangle \sim L_{\parallel}^{D_\parallel}$
where  $M_{i,n}$ is the mass of the $i$th largest spanning cluster
with $n$ spanning clusters present, $L_{\parallel}$ is the length parallel
to the preferred direction, $D_\parallel$ = ${-\beta /\nu_\parallel +1 +
(d-1)/b}$, $b = \nu_\parallel /\nu_\perp$,  $\nu_\parallel$
($\nu_\perp$) is the correlation length exponents 
parallel (perpendicular) to the
direction of the DP, and $\beta$ is the DP order parameter
exponent\cite{dprev,smb}. 
$D_\parallel = 1.47$ in two dimensions with the values of $\beta $,  
$\nu_\parallel$ and $\nu_\perp$ as given in \cite{jensen}. 
In the  $L \times L$ lattice with the direction 
of propagation  along the diagonal, it is expected that
$M_{i,n}  \sim L^{D_\parallel}$ as the spanning is considered
from top to bottom. This is because the length scale in the 
direction of propagation is $\sim L_\parallel$ and the width 
of the cluster is $\sim L_\perp$ and here $L_\parallel \sim L$.
We indeed find that $D \sim 1.5$ (comparable to
1.47)  here. See Fig. \ref{fig:mvsl}. For the DP parallel to the
spanning direction, $L_\parallel \sim L$  
obviously, and again the above scaling behavior is observed.
Only exception is the diagonal DP case with $\rho >1$
where  the mass exponent is  found to take a new value $\sim 1.8$. 
A value close to $2$ suggests a nearly compact structure
which could 
happen from strong boundary effect in this particular geometry.
The data of Fig \ref{fig:mvsl} show that $\langle 
M_{1,n}\rangle$ is independent of the number $n$ of clusters present,
signifying that the largest cluster is uncorrelated to the remainder
of the lattice.

The ratios of the masses of the individual spanning clusters 
should be constants  as the scaling behaviors are identical. Defining
$r_{i,j}^{\{n\}} = \langle M_{i,n}\rangle /\langle M_{j,n}\rangle$, we
find that  
$r_{1,2}^{\{2\}}$ is $1.36 \pm 0.03$ for the diagonal DP and $1.31 \pm
0.02$ for 
the parallel DP for $\rho =1$.  The variations of these ratios are
lesser when the aspect ratio is changed but the direction of the DP
remains same.  For comaprison, we add that 
$r_{1,2}^{\{2\}}\simeq 1.4$ for OP\cite{sen3}.

In three dimensional symmetric lattices ($L^3$), we find a significantly
different larger  number of spanning clusters compared to OP with 
the same number of random initial configurations ($10^6$).
However, when we plot $P(n)$ against $n$ and $n^2$, again the latter
gives a better straight line. The extrapolated value of   $\gamma_n$ is 
also possibly close to 2 (see Fig \ref{fig:pvsn2}). 
 The fractal dimension of the cluster from the known
values is around 1.67 and the log-log
plot of $M_{i,j}$ against $L$ is compatible with this value (Fig
\ref{fig:mvsl}. We obtain  
several mass ratios, for example,    
$r_{1,2}^{\{2\}} = 1.52 \pm 0.02,  
r_{1,2}^{\{3\}} = 1.34 \pm 0.03$,  
and $r_{1,3}^{\{3\}} = 1.77 \pm 0.05$. As $n$ becomes larger, the 
fluctuations also increase and the values of the 
ratios for $n \geq 4$ may be less reliable.
It maybe mentioned here that $r_{1,2}^{\{2\}} \simeq  1.8$ for OP. 
Though our results 
are not inconsistent with the expectation of angular
dependence in DP\cite{domany}, still it is perhaps premature to
comment quantitatively  
on the universality of these ratios from the present data.
However, one can safely conclude that at least for three dimensions,
these  ratios  are  not  same for OP and DP.

In order to probe the detailed 
mass distribution, we obtain
histogram for the cluster masses when DP is parallel to $y$ axis (the
TB-spanning direction). A collapse of all the probability
distributions is found, for various geometries and aspect ratios, if
a scaled variable $X=(M_{i}- \langle M_i\rangle )/S_i$ is used where
$S_i^2 = \langle M_i^2\rangle- \langle M_i\rangle^2$, and $M_i$ is the
$ith$ largest cluster irrespective of the value of $n$. No
sensitivity to $n$ was detected. 
We see 
a definite {\it lower} cutoff for the scaled variable.  The plot is
shown in Fig \ref{fig:pmvsx} which also shows a Pearson distribution
(type III).  Since by construction the scaled variable $X$ has a zero
mean and unit standard deviation, only parameter available for the
Pearson distribution is the lower limit.  The resulting curve for 
$-\alpha\leq x\leq \infty$,
\begin{equation}
Q(x)=\frac{\alpha}{\Gamma(\alpha^2)} [\alpha (x +\alpha)]^{\alpha^2
-1} e^{-\alpha(x +\alpha)},\label{eq:qx}
\end{equation}
with $\alpha=2.8$, fits remarkably the probability distribution
obtained numerically.

The single variable distribution $Q(x)$ does not reflect the
restriction $M_1 > M_2$.  For a joint distribution, we
study the probability density ${\cal R}(y)$ for the mass
ratio $y=M_1/M_2$ of the two largest clusters, again irrespective of
$n\geq 2$.  A data collapse with some finite size effect is seen, see
Fig \ref{fig:pmvsx}b.  An exponential distribution ${\cal R}(y)= A
\exp[-A(x-1)]$ with $A=4.$ is also shown there for comparison.  With
this exponential distribution, the average mass ratio is expected to
be $1 + A^{-1} \approx 1.25$.

Our main results have already been summarized at the beginning.  We
like to add a few comments here. 
First, the occurrence  of the two simple
distributions in Fig. \ref{fig:pmvsx} remains unexplained. The second
is regarding 
$\alpha$ in the distribution $P(n) \sim \exp(-\alpha n^2)$. 
As mentioned earlier, the number of TB-spanning
clusters at the threshold is infinity at the upper critical dimension
($d_c = 5$ for DP and $6$ for OP).  
validity for all $d<5$  for DP also. From $P(n)$,
Generalizing our result to all $d$ (as known to be the case for OP),
we may take  $n_c =  1/\sqrt{\alpha}$ as a measure of the number of
clusters. This
$n_c$, defined strictly at the critical point,  has to diverge as $d
\rightarrow d_c$.   Therefore,
$\alpha$ must decrease\cite{sen2} with dimension, vanishing at $d=d_c$.
We indeed find that $\alpha$ for DP is lesser than OP (for 
three dimensions,  where we could make a comparison), a fact
compatible with  $d_c$(DP)$<d_c$(OP). Assuming analyticity in
$\epsilon=d_c -d$,  one expects 
$\alpha=  a \epsilon +  b\epsilon^2+.. $, for small
$\epsilon$, with  geometry (and OP/DP) dependent $a$ and $b$.
In fact, for OP, 
 $a= 0.15$ and
$b = 0.05$ give very good agreement with the  
$\alpha $ values obtained for $\epsilon$ = 1,2 and 3 \cite{sen2}.
We conjecture that critical fluctuations are responsible for nonvanishing of 
$\alpha$, and therefore $\alpha$ is a new weakly universal
critical quantity.\\

We thank D. Stauffer for a critical reading of the manuscript.
PS is grateful to Saha Institute of Nuclear Physics for the
use of computer facilities.

\pagebreak

\end{multicols}

\begin{references}
\bibitem{exponent} Mass $M \sim L^D$ where $D = d- \beta /\nu$ is the 
usual size exponent, $\beta$ and $\nu$ being the
order parameter and length scale exponents (see,
e.g., Ref. \cite{staufferb}).
\end{references}
\end{document}